# Gut Microbiota-derived Bile Acids Promote Gamma-secretase Activity Through Interactions with Nicastrin Subunits

Hemi Luan[1#], Xuan Li[2#], Liang-Feng Liu[3*], Min Lin[3], Wenyong Zhang[2], Tiangang Luan[1,4,5*]

1.Department of Biomedical Engineering, School of Biomedical and Pharmaceutical Sciences, Guangdong University of Technology, Guangzhou, 510006, Guangdong, China

2. School of Medicine, Southern University of Science and Technology, Shenzhen, China

3. Mr. & Mrs. Ko Chi-Ming Centre for Parkinson's Disease Research, School of Chinese Medicine, Hong Kong Baptist University, Hong Kong SAR, China;

4. Institute of Environmental and Ecological Engineering, Guangdong University of Technology, Guangzhou 510006, China.

5. School of Life Sciences, Sun Yat-Sen University, Guangzhou 510275, China.

# These authors contributed equally to this work.

***Correspondence**

Prof. Tiangang Luan

Email : cesltg@mail.sysu.edu.cn

School of Life Sciences, Sun Yat-Sen University

**Abstract**

Alzheimer's disease (AD) has emerged as a progressively pervasive neurodegenerative disorder worldwide. Bile acids, synthesized in the liver and modified by the gut microbiota, play pivotal roles in diverse physiological processes, and their dysregulation in individuals with AD has been well-documented. However, the protein targets associated with microbiota-derived bile acids in AD have received limited attention. To address this gap, we conducted comprehensive thermal proteomic analyses to unravel and comprehend the protein targets affected by microbiota-derived bile acids in AD. Our investigation identified sixty-five unique proteins as potential targets of deoxycholic acid (DCA), a primary component of the bile acid pool originating from the gut microbiota. Particularly noteworthy among these proteins were Nicastrin and Casein kinase 1 epsilon. We found that DCA, through its interaction with the Nicastrin subunit of γ-secretase, significantly contributed to the formation of amyloid beta, a key hallmark of AD pathology. Additionally, We observed substantial elevations in the urine levels of four bile acids (DCA, GHCA, GHDCA, and GUDCA) in AD patients compared to healthy controls. Moreover, the ratios of DCA to cholic acid (CA) and glycodeoxycholic acid (GDCA) to DCA were significantly increased in AD patients, indicating aberrations in the biosynthetic pathway responsible for bile acid dehydroxylation. The augmented levels of microbiota-derived bile acids and their altered ratios to primary bile acids exhibited notable associations with AD. Collectively, our findings provide crucial insights into the intricate interplay between microbiota-derived bile acids and the pathogenesis of AD, thereby shedding light on potential therapeutic targets for this debilitating disease.

## Introduction

Alzheimer's disease (AD) is a progressive neurodegenerative disease that affects millions of people worldwide. The prevalence of AD is increasing, and it is estimated that by 2050, the number of cases will triple(1). The cause of AD is not fully understood, but it is believed to be a combination of genetic and environmental factors, including the gut microbiome (2, 3). The gut microbiota is a complex ecosystem of microorganisms that reside in the gastrointestinal tract. It plays a crucial role in human health, including digestion, immune system regulation, and metabolism. The gut microbiota also produces metabolites including vitamins, secondary bile acids, and neurotransmitters that may contribute to the development of neurodegenerative diseases (4-6).

Bile acids are produced in the liver and play a crucial role in lipid digestion and absorption. Bile acids also act as signaling molecules that regulate energy metabolism, glucose homeostasis, and inflammation (7, 8). The gut microbiota further modifies the primary bile acids by deconjugation, dehydroxylation, and epimerization, resulting in the formation of secondary bile acids. These secondary bile acids have been shown to modulate various physiological processes, including inflammation, glucose metabolism, and gut permeability (9). Recent works show dysregulation of bile acids has been observed in the cerebrospinal fluid of AD patients (10). Increased levels of microbiota-derived bile acids (MBA) are associated with AD and poor cognitive performance. MBA may influence AD pathogenesis by modulating the blood-brain barrier (BBB). The BBB is a critical barrier that separates the brain from the circulatory system and regulates the entry of substances into the brain. The study found that MBA could modulate BBB permeability and promote the entry of neurotoxic substances into the brain, leading to neuroinflammation and neuronal damage (11). MBA may modulate AD pathogenesis by influencing Aβ aggregation by altering the gut microbiota composition. Aβ is a peptide that aggregates to form amyloid plaques, a hallmark of AD pathology (12).

Multiple proteomic technologies have been developed for the comprehensive analysis of entire proteomes, revealing hitherto unknown interactions between MBA and proteins (13-15). The traditional molecular biotechnologies have identified the farnesoid X receptor (FXR) and Takeda G protein-coupled receptor 5 (TGR5) as two well-characterized BA receptors. However, little attention has been paid to the protein targets of MBA connected with Alzheimer's disease. Thermal proteome profiling (TPP) is a relatively new and powerful tool in the field of proteomics that allows for the identification of protein-ligand interactions in cell lysates, tissues, or living cells (16). The technique involves exposing cells to different temperatures, which can cause proteins to undergo conformational changes or denaturation. These changes can be detected using quantitative mass spectrometry, allowing us to identify and quantify proteins that interact with MBA under different thermal conditions (17). In light of this, we have perform comprehensive thermal proteomic analyses to unravel and comprehend the protein targets affected by microbiota-derived bile acids in AD, intending to understand the complex interplay between MBA and AD.

## Result and discussion

The workflow (Fig. 1) described the discovery of novel protein targets of MBA and helps to reveal the mechanisms of interaction with proteins in cellular metabolism of active metabolites. A secondary bile acid DCA derived from the gut microbiota interacts with the Nicastrin subunit from γ-secretase and contributes to the formation of Aβ.

Figure 1

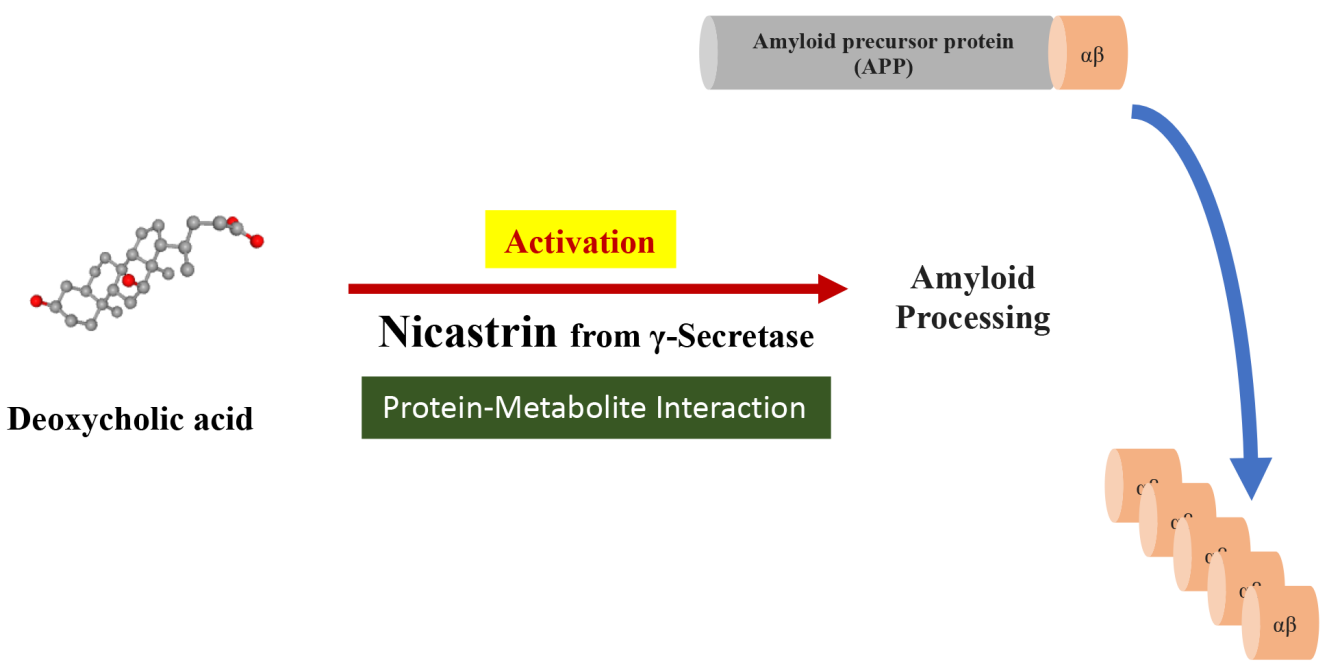

***Figure 1****. A secondary bile acid DCA derived from the gut microbiota interacts with the Nicastrin subunit from γ-secretase and contributes to the formation of Aβ.*

## TPP for identification of protein targets of DCA

TPP has been widely used to measure protein thermal stability and abundance for unbiased identification of direct and indirect drug targets (16). It offers a number of advantages over traditional proteomics approaches, including high throughput, sensitivity, precision, accuracy, and the ability to provide insights into complex biological systems (18). We analyzed the proteomic changes after incubation of live cells with DCA. Theoretically, the heat stability of the target proteins can be enhanced upon ligand binding in solution than the unbound proteins (19). Therefore, these DCA-bound proteins would be more abundant in the supernatant after DCA treatment than in the control after thermal challenge and centrifugation. We evaluated the melting curve of SH-SY5Y lysate after the incubation with DCA at conditions ranging from 37 to 67°C and found that the median protein thermal melting temperature (Tm) was 53°C, which was subsequently selected as the temperature for screening DCA target proteins. Fig. 2A shows the detailed workflow of TPP applied to screen target proteins for DCA in living cells. Briefly, cells were treated with DCA and DMSO and heated at 37 and 53 °C for 3 mins. Subsequently, proteins were extracted, digested, and further labeled with iTRAQ stable isotope reagents for quantitation analysis.

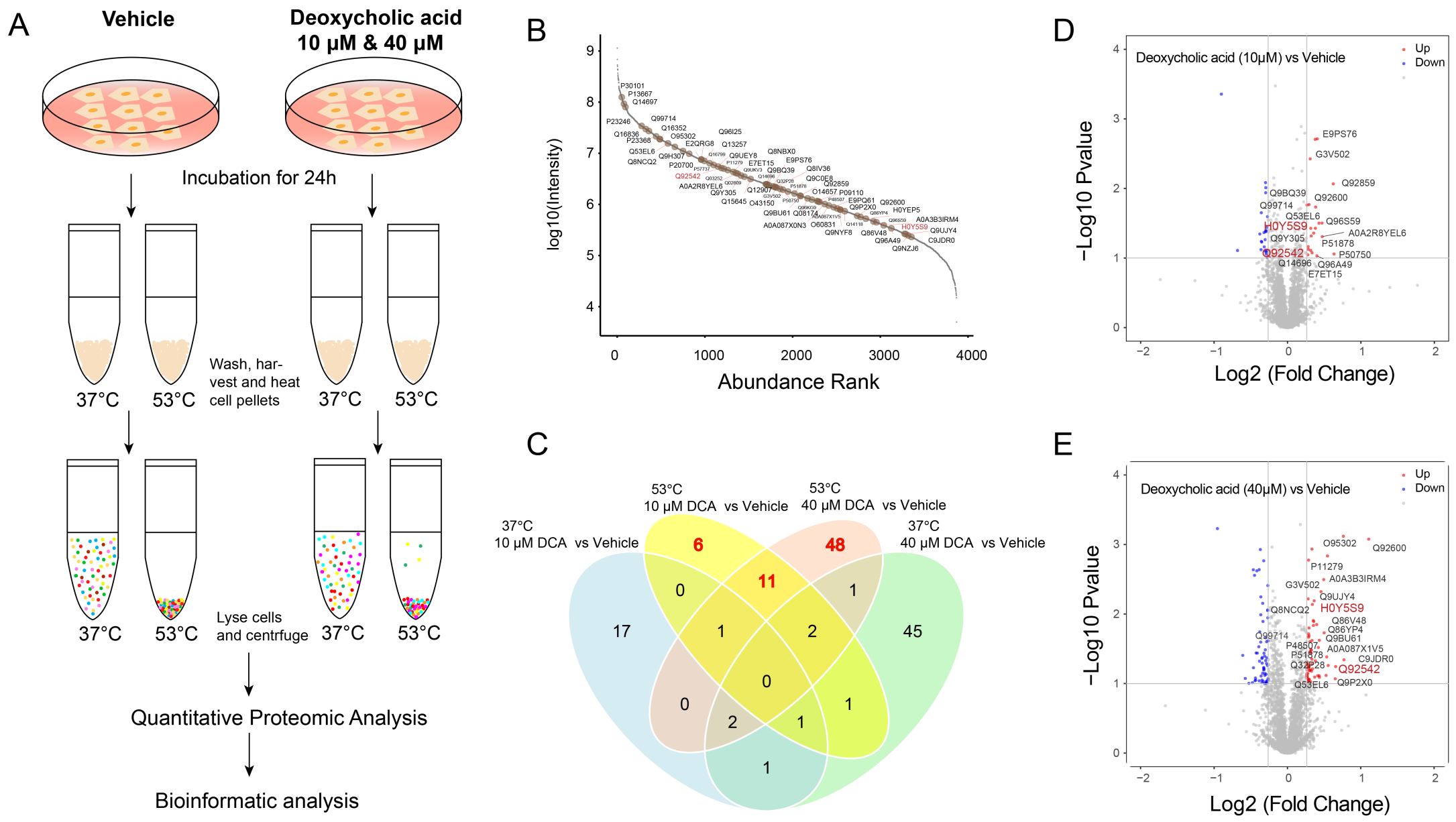

***Figure 2.*** *TPP for identification of protein targets of DCA.* ***A****, Workflow of TPP experiments.* ***B****, Abundance rank plot of identified 4076 proteins.* ***C****, Venn plot showing stabilized protein targets shared by different groups.* ***D and E****, Volcano plot showing protein targets stabilized in cells following administration of DCA (10 μM or 40μM).*

As shown in Fig. 2B, A total of 4076 protein groups were annotated from 22679 peptide groups. After normalization of the quantitative data with the vehicle, proteins with significant enrichment fold change not less than 1.2 and adjusted p-value less than 0.05 in the 53 °C heated DCA treatment groups were selected as the potential targets. Overall, we have identified 65 unique proteins as the potential targets of DCA in the 10 μM and 40 μM treatment groups under 53 °C (Fig. 3C, D, and E). Among them, Nicastrin (NCSTN) and Casein kinase 1 epsilon (CSNK1E) were identified with high confidence in the DCA treatment groups, compared with the vehicle group (Fig. 2D and E).

Although the use of chemo-proteomic probes has led to the identification of hundreds of bile acid-interacting proteins, encompassing recognized receptors, transporters, and biosynthetic enzymes of BA (20), It is worth noting that the NCSTN and CSNK1E are novel BA–protein interactions discovered in our study. NCSTN is a highly glycosylated type I transmembrane protein and is widely distributed in all cell types of humans or mice as one of the important component proteins of the γ-secretase complex. It is not only closely related to the assembly and maturation of γ-secretase but also plays an important role in regulating the production and degradation of γ-secretase activity and β-amyloid protein in AD (21). Another MBA-interacting protein, CSNK1E, is part of the CK1 family of ubiquitous serine/threonine-specific protein kinases, which has been suggested to have a role in AD pathology. CSNK1E is involved in regulating the cleavage of the γ-secretase of APP, leading to increased formation of Aβ peptides (22, 23). Our study identified NCSTN and CSNK1E as MBA-interacting proteins and may provide a novel link for MBA in the direct regulation of Aβ peptide formation.

**MBA-interacting proteins and pathway enrichment analysis**

Screening of MBA-interacting proteins from cells treated with different drug concentrations and temperatures was employed to identify bile acid-interacting proteins. Overall, 65 out of 4076 proteins were identified as MBA-interacting proteins, which has their good thermal stability upon interactions with DCA (Fig. 3A). The 65 MBA-interacting proteins were analyzed through the use of Ingenuity

Pathway Analysis (IPA, QIAGEN INC.) to identify enriched pathways and associated networks based on a calculated probability score of ≥2 and a p-value of <0.05 using IPA software application. The top-ranked canonical pathways influenced by these 65 proteins were shown in Fig. 3B. The pathways associated with "Glutathione Biosynthesis ", and "Glutathione Redox Reaction II" were the most significantly enriched pathways, indicating MBA may be associated with the antioxidant defense mechanism and oxidative damage (24). Notably, two MBA-interacting proteins (NCSTN and CSNK1E) were highlighted in the amyloid processing pathway (Fig. 3B). From the data for canonical pathways and biological functions, IPA assessed networks which revealed the direct or indirect association of the 65 enriched MBA-interacting proteins with each other. The first and most relevant network is depicted in Fig. 3C. This network included 26 of 65 the analyzed MBA-interacting proteins. A central component of this network includes γ-secretase and NCSTN, which are associated with the processing of amyloid precursors in Alzheimer's disease (25).

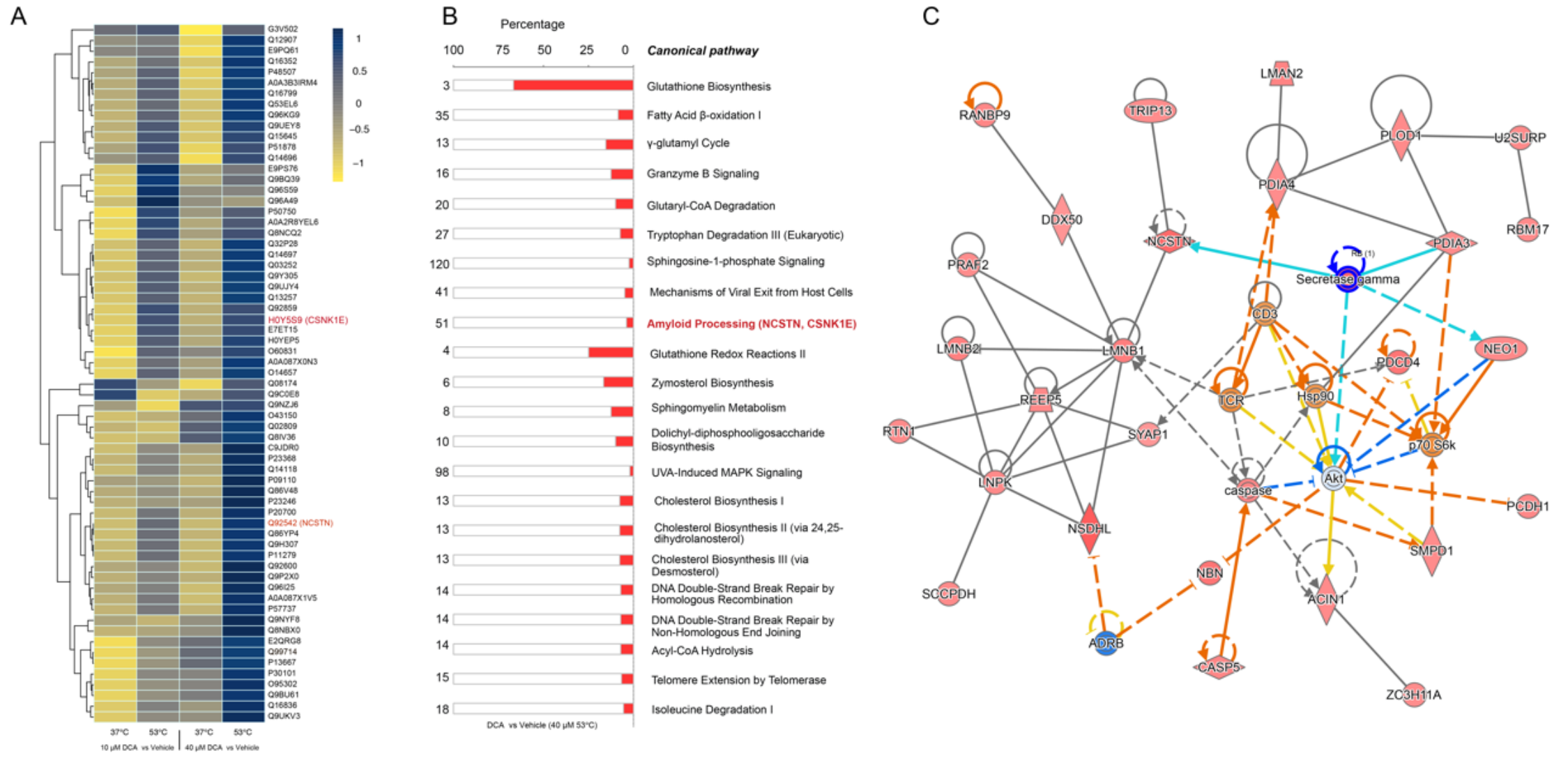

***Figure 3.*** *DCA-interacting proteins and pathway enrichment analysis.* ***A****, The levels of stabilized protein targets were visualized by the depicted heat map.* ***B****, The top-ranked canonical pathways.* ***C****, typical protein-protein interactions networks revealed the association of stabilized protein targets.*

**Binding affinity evaluation of DCA and protein targets**

We generated isothermal (37 °C and 53 °C) dose-response curves and calculated $EC_{50}$ values to evaluate the binding affinities between DCA and the involved MBA-interacting proteins. The cell lysates were treated with two different doses of DCA as well as one vehicle control. As shown in Fig. 4A, an increase in the concentration of the target proteins was observed as the ligand concentration increased to a level of protein binding saturation. The $EC_{50}$ values of NCSTN and CSNK1E were calculated as 6.03 μM and 7.22 μM, respectively, indicating the binding affinity between DCA and target proteins (26).

We further evaluated the binding affinity between DCA and the involved proteins (NCSTN and CSNK1E) by molecular docking analysis (27). Figure 4B shows that DCA could be embedded into the active pocket of NCSTN and CSNK1E with the binding energy of -7.9 and -7.8 kcal/mol, respectively. DCA is capable of hydrogen-bonding, salt-bridging, and hydrophobic interactions with the binding pocket of NCSTN protein. Subsequently, by analyzing the three-dimensional interactions, it was found that the DCA is capable of forming hydrogen-bonding interactions with proteins TYR283, THR387, ASP447, and ILE468; the carboxylic acid functional group of the compounds is negatively charged under physiological pH conditions, and the negatively charged functional group of the compounds forms salt-bridging interactions with the positively charged LYS470 of the protein. Similarly, DCA is also capable of hydrogen-bonding and hydrophobic interactions with the CSNK1E protein binding pocket and hydrogen-bonding interactions with proteins ALA56, THR57, and ASP336. The compound is capable of hydrophobic interactions with ASP143 and TYR173 of the protein. These interactions promote DCA binding to the active pocket of the protein to form a complex. These results indicated that NCSTN and CSNK1E are the targets of DCA, and the binding strength is positively correlated with the concentration of DCA.

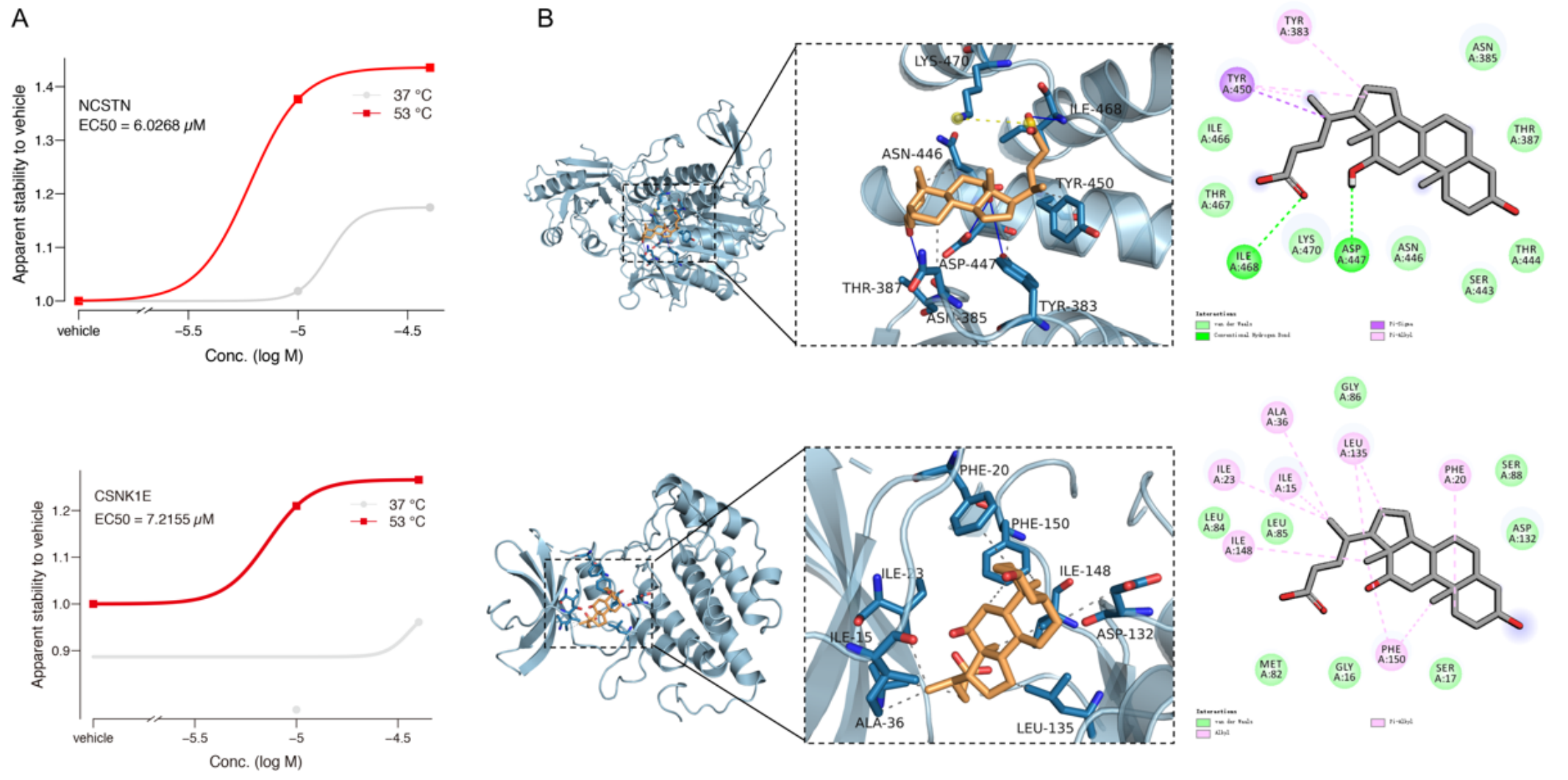

***Figure 4.*** *Binding affinity evaluation of DCA and protein targets.* ***A****, Dose-dependent stabilization of NCSTN and CSNK1E by DCA at 37°C and 53°C, respectively. Stabilization of NCSTN and CSNK1E in response to DCA treatment was measured relative to vehicle-treated control and fitted with saturation curves*

*to determine the concentrations leading to half-maximal stabilization, $EC_{50}$ (−log10-transformed compound concentrations).* ***B****, Evaluation of the binding affinity between DCA and the involved proteins (NCSTN at the top layer and CSNK1E at the bottom layer) by molecular docking analysis.*

**DCA can interact directly with NCSTN and affect the formation of Aβ**

NCSTN is the putative substrate recruitment component of the γ-secretase complex that is responsible for the production of Aβ peptides (25). We further evaluated whether DCA can affect the formation of amyloid β. We have observed that 10 and 40μM DCA treatments have resulted in the enrichment of NCSTN and CSNK1E playing vital roles in the amyloid processing at 53 °C, indicating that the processing of amyloid precursor protein to Aβ might be an important target pathway of DCA (Fig. 4A and 5A). In contrast, there was no significant change in the concentration of these proteins after 37 °C heating in DCA-incubated samples, suggesting that the DCA does not affect their constitutive abundance (Fig. 4A and B). Furthermore, we incubated different concentrations of DCA with live cells. As shown in Fig. 5C and D, the concentration of Aβ40 in live cells was significantly increased under 10 and 40 μM DCA treatment. We also found that APP protein was significantly increased only under high concentrations of DCA treatment. These results suggest that DCA may promote the conversion of amyloid precursor protein to Aβ.

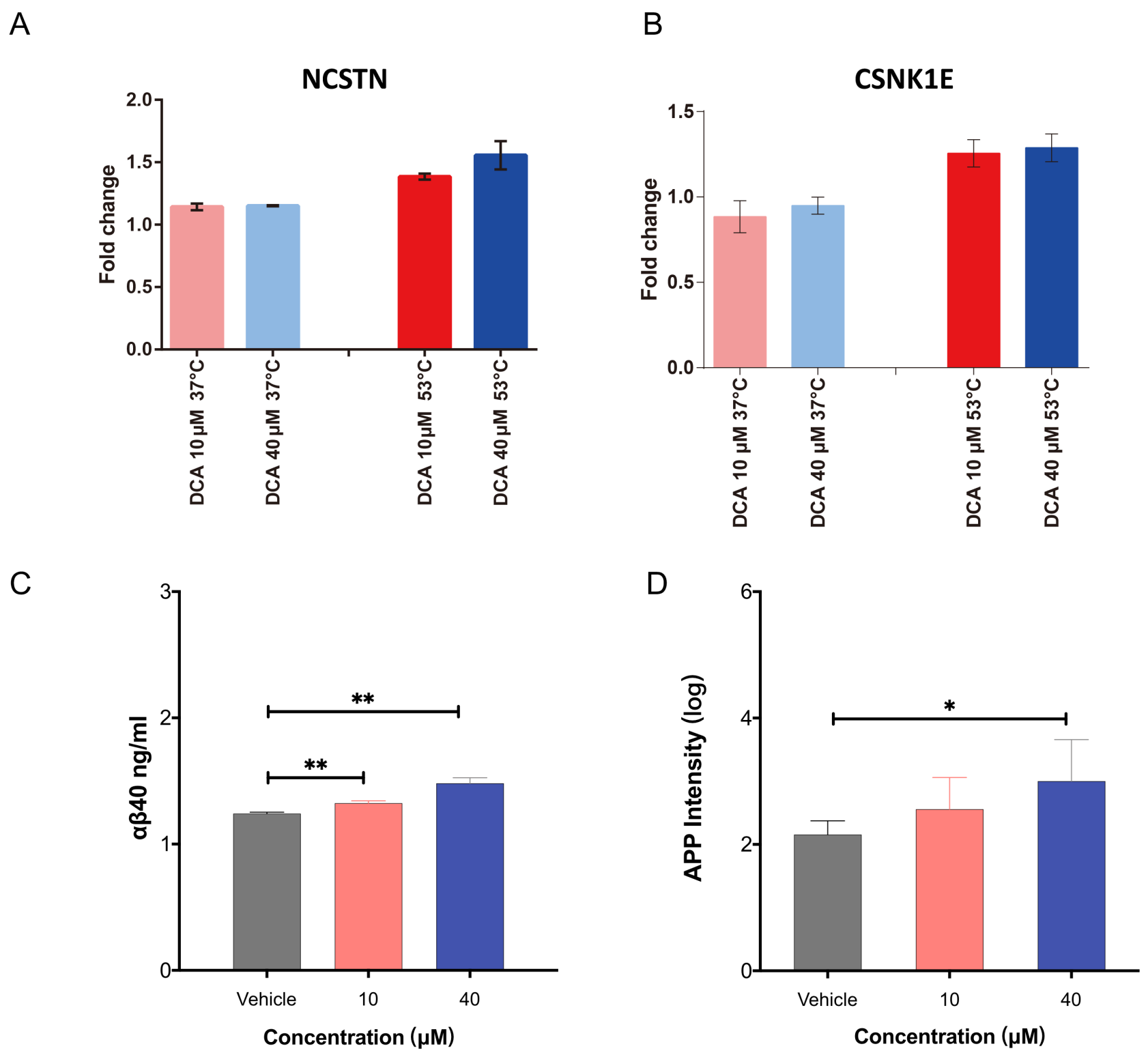

***Figure 5.**. DCA interacts with NCSTN and affect the formation of Aβ. **A and B**, The level changes of NCSTN and CSNK1E. The column marked as red and blue refers to the cells treated with DCA 10 µM at 37 and 53 °C, respectively, (data are shown as mean ± SE, n = 3). The DMSO treatment groups (37 or 53 °C) were normalized as 1, while the fold changes of DCA treatment groups were calculated after normalization with the vehicle group. **C and D**, bar plots showing increased αβ 40 and increased Amyloid precursor protein (APP) levels upon DCA treatment. Asterisk (*) indicates a statistically significant difference ($p < 0.05$) between DCA-treated and vehicle (DMSO) groups.*

## Association of MBA accumulation with Alzheimer's disease

MBA are produced by the gut microbiota and play a crucial role in the regulation of host metabolism (28, 29). There has been growing interest in the potential role of MBA in the pathogenesis of AD (30-32). In the current study, non-invasive metabolomics was employed to quantify 15 urinary bile acids of 83 normal control subjects and 36 AD patients. We found 4 bile acids such as DCA, GHCA, GHDCA, and GUDCA were significantly increased in the urine of AD patients compared with normal control subjects, which is in line with previous findings of perturbation in bile acids in invasive serum metabolome of AD patients compared with healthy controls (30). The levels of urinary primary bile acids including CA and CDCA have no significant differences between AD patients and normal control subjects (Fig. 6A). We found the ratio of DCA to CA and GDCA to DCA were significantly increased in the urine of AD patients compared with normal control subjects, indicating the aberration in the biosynthetic pathway for bile acid dehydroxylation (Fig. 6B and C). DCA is one of the most abundant secondary bile acids, which are produced by cholic acid catalyzed by 7α-dehydroxylase in the intestinal flora, leading to neuroinflammation and neuronal death (33, 34). Our results highlighted urinary MBA can noninvasively reflect changes in the status of the human bile acid pool, and DCA accumulation was associated with AD development.

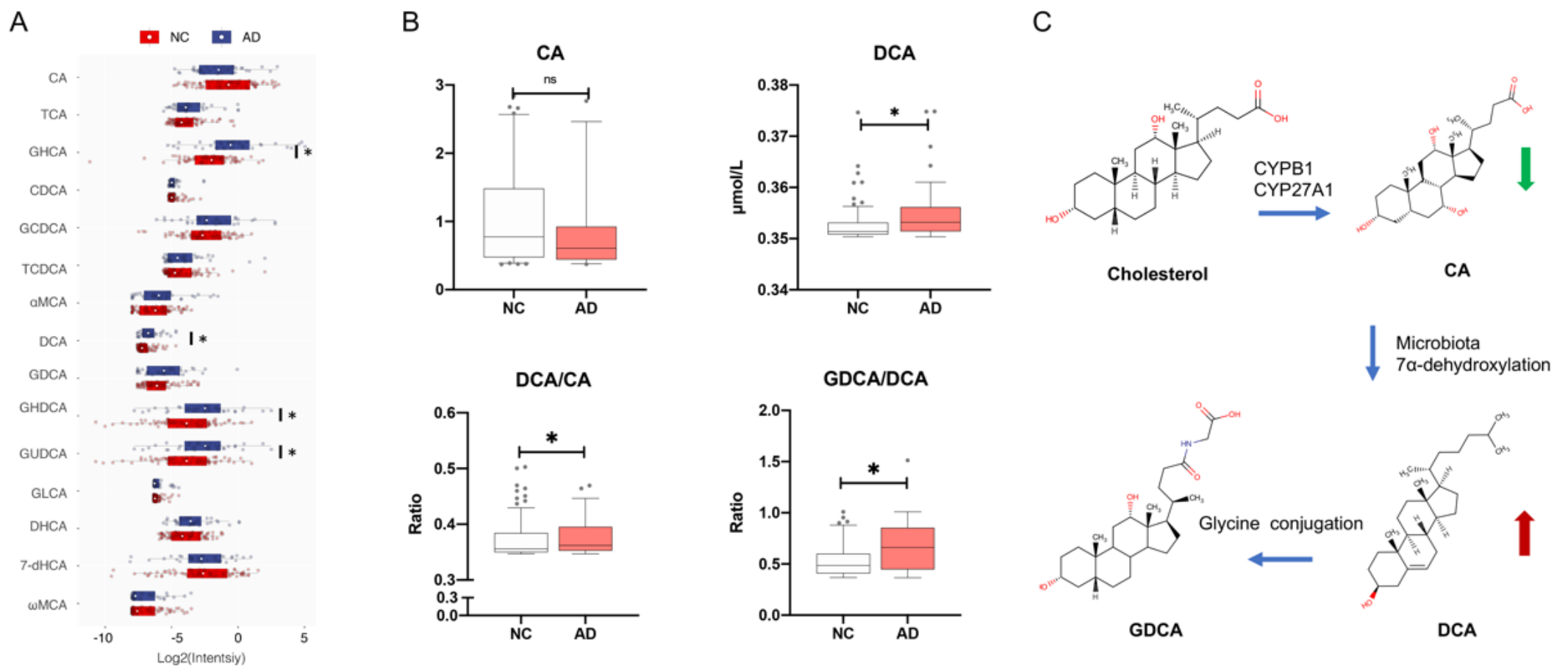

***Figure 6.**. Association of MBA accumulation with Alzheimer's disease. **A**, The levels of urinary bile acids were visualized by the depicted heat map. **B**, boxplot showing significantly changed bile acids in AD patients. **C**, the aberration in the biosynthetic pathway for bile acid dehydroxylation. DCA, Deoxycholic acid; CA, Cholic acid; GDCA, Glycodeoxycholic acid; AD, Alzheimer's disease patients; NC, Normal control subjects. Asterisks (*) indicate statistical significance ($p < 0.05$)*

## Conclusion

The intricate interplay between gut microbiota and host metabolism has been the subject of numerous investigations (4). Of particular interest are microbial bioactive metabolites, such as bile acids, amino acids, and short-chain fatty acids, which have been implicated in neurodevelopmental processes and neurodegenerative disorders (28, 35, 36). Despite this knowledge, the specific protein targets of MBA remain elusive. In our study, employing the TPP method, we successfully identified NSCTN and CSNK1E as direct protein targets of deoxycholic acid (DCA), an MBA associated with Alzheimer's disease. Furthermore, we confirmed the ability of DCA to promote the formation of Aβ, a hallmark of Alzheimer's pathology. Additionally, by examining the metabolic conversion of DCA in urine, we established its potential as a risk indicator for Alzheimer's disease. Our findings shed light on the underlying mechanisms connecting MBA and neurodegenerative diseases, providing valuable insights into these complex pathological processes.

## Materials and methods

### Reagents and cell culture

Deoxycholic acid (DCA), Cholic acid (CA), Glycodeoxycholic acid (GDCA), Taurocholic acid (TCA), Glycohyocholic acid (GHCA), Chenodeoxycholic acid (CDCA), Taurochenodeoxycholic acid (TCDCA), Muricholic acids (MCA), Glycohyodeoxycholic acid (GHDCA), Glycolithocholic acid (GLCA), Dehydrocholic acid (DHCA), Deoxycholic acid-d4, Dimethyl sulfoxide (DMSO), Dithiothreitol (DTT), Iodoacetamide (IAA) were purchased from Sigma-Aldrich with purity for analytical standards (>99%). Isopropyl alcohol (IPA), acetonitrile (ACN), Trypsin enzymes and iTRAQ® Reagent Application Kit were purchased from Sigma-Aldrich. A bicinchoninic acid (BCA) protein assay kit were purchased from Thermo Scientific. All solvents used in this study were high-performance liquid chromatography grade or higher. SH-SY5Y cells lines were obtained from the America Type Culture Collection (ATCC).

### Human participants

A total of 36 AD patients and 83 normal control subjects were recruited at the Hong Kong Baptist University Chinese Medicine Specialty Centre. The study was approved by the Ethics Committee of the Hong Kong Baptist University's Institutional Review Board with written information and informed consent obtained from all subjects. Urinary samples were collected from all subjects using the same protocol as follows: After overnight fasting, urine was collected and stored at −80 °C for analysis.

**Bile acids extraction**

An aliquot of 100 µL urine sample was mixed with 100 µL acetonitrile–methanol (8:2, v/v). The mixture was centrifuged at 20000 × g at 4 °C for 15 min. The supernatant was evaporated under vacuum at room temperature and reconstituted in acetonitrile–methanol (9:1, v/v) containing 0.01% formic acid. The sample was vortexed and then centrifuged at 20000 × g, 4 °C for 15 min. The supernatant was used for LC-MS/MS analysis.

**Metabolomic analysis**

The metabolomics analysis of bile acids was conducted using a TSQ Quantiva™ triple quadruple mass spectrometer (Thermo Fisher Scientific, MA, USA) equipped with an electrospray ionization (ESI) source. LC separation was performed using a BEH C18 (2.1 mm*100 mm, 1.7 µm) column with an injection volume of 5 µL. The mobile phase was consisted with 0.01% formic acid in water and 0.01% formic acid in acetonitrile. The chromatographic separation was achieved using gradient elution with a total run time of 15 min. The flow rate was constant at 350 µL/min. The autosampler and column oven temperatures were maintained at 10 C° and 50 C°, respectively. Multiple reaction monitoring (MRM) mode was used for detection of bile acids. The negative ion electrospray voltage was 3500 V, and the optimal gas flows were 40, 10, and 0 for Sheath, Aux, and Sweep Gas, respectively. The calibration curve was built by using the IS-normalized peak area ratio and the concentrations of standards.

**Sample preparation for TPP**

SH-SY5Y cells cultured in the 10 cm dishes were treated with 0, 10 and 40 µM DCA in a 5% CO2 incubator at 37°C for 24h respectively. DCA was dissolved in DMSO and final concentration of DMSO was 0.1%. Then cells were washed with 3 mL cold PBS twice and scraped down with another 1 mL cold PBS supplemented with EDTA-free protease inhibitor cocktail. After centrifugation at 300 g for 5 min at 4°C, the cell pellets were resuspended in 200 µL PBS and transfered equally to new PCR tubes. Subsequently the samples were heat at 37 or 53°C for 3 min and followed by incubation at room temperature for 3 min. Afterwards, cells were lysed by repeated freezing and thawing and centrifuged at 20000 × g for 20 min at 4°C. Then supernatant was collected and BCA assay was performed to determine the protein concentration.

**Tryptic digestion, ITRAQ labeling**

The same amount of proteins from each sample were reduced by 10 mM DTT for 30 min at 55°C and then alkylated with 30 mM IAA for 30 min at room temperature. Excess IAA was quenched by 10 mM of DTT. Proteins were then purified by acetone precipitation and protein precipitates were dissolved in 50 μL Dissolution buffer from the iTRAQ reagent kit (0.5M TEAB, PH8.5). Proteins were digested overnight by trypsin at a ratio of 20:1 (protein:trypsin). After digestion, each peptide sample was incubated with iTRAQ reagent at room temperature for 2 hours. HPLC grade water was added to quench the reaction. Then the combined iTRAQ-labeled peptides were desalted and fractionated on a SCX and C18 column using stepwise elution with 3%, 6%, 9%, 15% and 80% (v/v) ACN in 0.1% ammonia. Each fraction was lyophilized and resuspended in 0.1% (v/v) formic acid for MS analysis.

**LC-MS/MS analysis**

LC-MS/MS analysis was performed on an Easy-nLC 1000 (Thermo Fisher Scientific) chromatography system coupled to an Orbitrap Fusion mass spectrometer (Thermo Fisher Scientific). Peptide samples were separated on an analytical column (75 μm i.d. × 15 cm, packed with 3 μm C18 particles) using a gradient of 50 min from 5 to 18% phase B (0.1% FA in acetonitrile), followed by 10 min from 18 to 28% at a flow rate of 250 nL/min. Survey full scan MS spectra (m/z 350-1550) were acquired in the Orbitrap with a resolution of 120,000, AGC target value of $2\times10^6$, and maximum injection time of 40 ms. Data-dependent acquisition of MS/MS spectra was performed using a top speed of 3 seconds and dynamic exclusion of 30 seconds. Precursors were selected using an isolation window of 1 Da and then fragmented by HCD using collision energy of 38%. Fragment ions were recorded by the orbitrap mass analyzer with first mass of 100 m/z, orbitrap resolution of 30,000, AGC target value of $5\times10^4$, and a maximum injection time of 60 ms.

**Protein identification and quantification**

Raw data were searched through UniProt Homo sapiens proteome database via Proteome Discoverer 2.4 software (Thermo Scientific). The search results were filtered to 1% false discovery rate to reduce the probability of false peptide identification. iTRAQ quantitation was performed according to previously reported rules.

**Pathway analysis and functional annotation**

The functional annotation and pathway enrichment analysis were integrated and analyzed in Ingenuity Pathway Analysis (IPA, QIAGEN INC.).

**Molecular Docking**

DCA and protein structure files were downloaded from the PUBCHEM database and Protein Data Bank, respectively, as described for each docking software program. Autodock vina v1.1.2 was used to visualize ligand and receptor docking, and the final results were visualized by PYMOL and Discovery studio software.

**Statistical analyses**

The proteins with fold changes greater than 1.2 and p-value less than 0.05 in either 10 μM or 40 μM DCA-treated group under 53°C conditions while they remained unchanged at 37°C were screened as DCA's protein targets. Comparisons between two groups were performed using Welch's t-test. The relative protein ratio was plotted versus DCA concentration and fit to a standard four-parameter logistic model to calculate $EC_{50}$. Heatmap and volcano plot were generated in R package "pheatmap" and "statTarget" (37), respectively.

## Acknowledgments

The authors would like to acknowledge the financial support from the National Natural Science Foundation of China (Grant No. 22376034).

## Conflict of Interest

The authors declare no competing interests.